\documentclass[10pt,twocolumn]{article} 
\usepackage{simpleConference}
\usepackage{times}
\usepackage{graphicx}
\usepackage{amssymb}
\usepackage{url,hyperref}

\begin{document}


\title{Shared-Protected Backup Paths Assignment with Mode Group Division Multiplexing in Optical Networks}%


\author{Jiaheng Xiong$^{(1)}$,  Qiaolun Zhang$^{(1)^{*}}$, Ruikun Wang$^{(2)}$, Alberto Gatto$^{(1)}$, \\Francesco Musumeci$^{(1)}$, Massimo Tornatore$^{(1)}$\\
$^{(1)}$Politecnico di Milano, Corresponding author: $^{(*)}$qiaolun.zhang@mail.polimi.it  \\
$^{(2)}$Beijing University of Posts and Telecommunications, Beijing, China
}


\maketitle

\begin{abstract}
    We evaluate the resource efficiency of Mode Group Division Multiplexing (\textit{MGDM}) with shared path protection (SPP) in optical networks. On our case studies, SPP with \textit{MGDM} obtains significant savings in terms of both additional backup spectrum occupation and MIMO-computing resources compared to other few-mode-transmission scenarios.
\end{abstract}
    


\section{Introduction}

To accommodate the continuous traffic growth, innovative solutions to enhance the capacity and efficiency of optical networks are constantly needed. Few-mode transmission (FMT) using Mode Division Multiplexing (MDM) has emerged as a promising technology by transmitting multiple spatial modes over a single wavelength to increase channel capacity~\cite{gatto2024partial}. Mode Group Division Multiplexing (MGDM), a variant of FMT, is particularly interesting for its scalability and reduced cost. Specifically, \textit{MGDM} classifies different spacial modes into several mode groups (MGs), where the inter-group crosstalk is small enough for sufficient quality of transmission. 
The granularity of resource sharing in \textit{MGDM} is reduced to MGs rather than wavelengths, offering greater flexibility in resource allocation~\cite{boffi2021mode, rottondi2016routing}. 
Furthermore, \textit{MGDM} requires only partial Multiple Input Multiple Output (MIMO) processing to recover just the spatial groups belonging to the same MGs, which provides substantial bandwidth enhancements for high-demand applications like Beyond-5G/6G backhaul with reduced MIMO complexity~\cite{boffi2021mode, rottondi2016routing, munoz2019adaptive}.




Following the experimental studies on \textit{MGDM}\cite{tsekrekos2008mode, de2005first,Gatto2024}, some prior studies have already investigated resource allocation with FMT, but only a few works~\cite{zhang2023resource} have investigated network resiliency in optical networks with FMT.
In particular, to the best of our knowledge, no existing work has focused on using Shared Path Protection (SPP) with \textit{MGDM}. 
MGDM with SPP has the potential to maintain resource efficiency and reduce the need for extensive MIMO processing.  Specifically, the backup path might use different MGs (due to longer transmission distance and availability of spectrum resources) compared to the working path, requiring additional MIMO as the MIMO for different MGs are different, but these additional MIMO for backup paths can be shared among the backup paths of link-disjoint working lightpaths, leading to more sustainable backup resource requirements. 



This study aims to evaluate the resource efficiency of SPP using various FMT approaches. Specifically, we develop a heuristic algorithm to quantify the advantages of different FMT methods in terms of spectrum occupation and MIMO computing resources for the backup paths of SPP compared to Dedicated Path Protection (DPP).

\section{Classification of Single-Mode and Few-Mode Transmission Approaches}

\begin{figure*}[h] 
    \centering 
    \includegraphics[width=\textwidth]{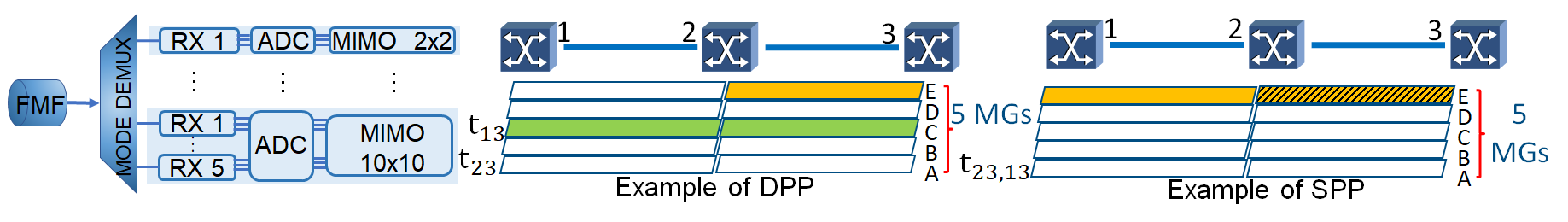} 
    \caption{(a) illustration of MGDM MIMO complexity, (b) Comparison between SPP and DPP.} 
    \label{fig:node-architectures} 
\end{figure*}

To evaluate the performance of SPP with \textit{MGDM}, we consider four transmission scenarios:
1) Single-mode transmission (\textit{SMT}) uses one mode, offering simplicity at the cost of limited capacity.
2) \textit{MGDM} transmission classifies the modes into five MGs, named A, B, C, D, and E, which contain 1, 2, 3, 4, and 5 modes, respectively~\cite{gatto2022mode}. 
3) MIMO-free MGDM (\textit{MF-MGDM}) reduces the MIMO complexity of MGDM by transmitting only one mode in each MG. However, this approach results in a smaller capacity and shorter transmission distance due to differential MG delay (DMGD)~\cite{gatto2022mode}. 
4) \textit{Full MIMO} transmission exploits all MGs for high data rates. 
Fig.~\ref{fig:node-architectures}(a) illustrates the architecture and the corresponding MIMOcomputing resources (defined as the number of equalizers~\cite{boffi2021mode} employed in the receiver DSP, normalized to those used in a standard single mode detection setup with a $2x2$ MIMO configuration) of each MG in \textit{MGDM}. FMF means Few Mode Fiber. As shown in Fig.~\ref{fig:node-architectures}(a), different MGs can only use partial MIMO. The MIMO complexity of a MG with $k$ modes can be calculated with $2k \times 2k$. 
We utilize the reach table as in Ref.~\cite{zhang2023resource} with multiple modulation formats. Note that, due to large inter-group crosstalk and DMGD, \textit{MGDM} and \textit{MF-MGDM} have a shorter transmission distance than \textit{Full MIMO}. 



\begin{figure*}[ht] 
    \centering 
    \includegraphics[width=0.93\textwidth]{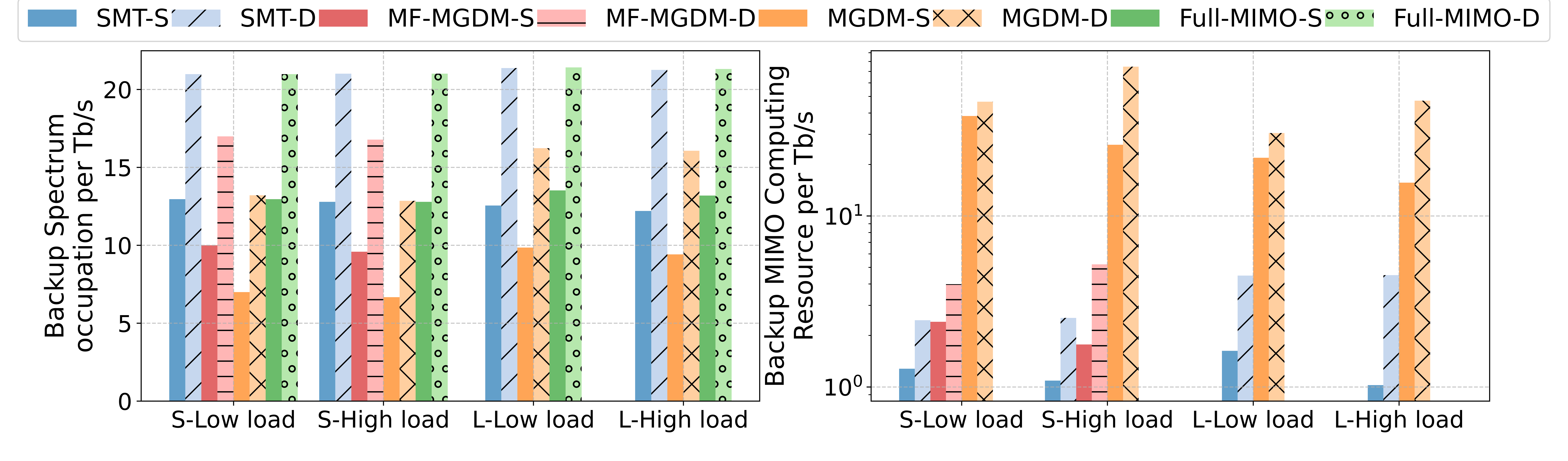} 
    \caption{(a) Average spectrum occupation of SPP and DPP, (b) Average MIMO complexity of SPP and DPP} 
    \label{fig:complexity and spectrum} 
\end{figure*}

\section{Shared Path Protection (SPP) and Dedicated Path Protection (DPP) with MGDM}

We define the resource-sharing principle of SPP with \textit{MGDM} as follows. If the backup paths of two requests traverse the same physical link, these two backup paths can share the same MGs on the same wavelength in the shared physical link as long as the corresponding primary paths of these two backup paths do not traverse the same physical link. For MIMO computing resources, the backup path can reuse receivers in working path to reduce the MIMO complexity. But, if additional MIMO, beyond what is available on the working path, is required for the backup paths, this MIMO can be shared among different requests as long as their working paths are link-disjoint (as for the case of physical link capacity sharing).

Fig.~\ref{fig:node-architectures}(b) shows an example of SPP and DPP with \textit{MGDM}, where
we have two backup requests, $(1,3)$ and $(2, 3)$ (denoted with $t_{1,3}$ and $t_{2,3}$, respectively) to be routed over any of 5 possible MGs over a given wavelength (the primary paths for these requests are not depicted in the figure as they are not the focus of this discussion, but we assume that the primary paths for these two requests do not overlap with each other). Different colors indicate different MGs (on the same wavelength), and the sharing of a mode group on a specific link is denoted by dashed lines. In the DPP scenario, requests $t_{1,3}$ and $t_{2,3}$ can not use the same MGs, and are served with MG A and C, respectively. 
Instead, for SPP, request $t_{1,3}$ can share the same MG A with $t_{2,3}$, and the MG C is not occupied. Let us now assume in this example that the MG assigned to the backup paths (i.e., MG E) is different from the MG assigned to the working paths (e.g., MG A). Hence, as MG E requires larger MIMO computing resources than MG A, we must deploy additional MIMO computing resources for the backup paths, but these additional backup MIMO resources in node 3 can now be shared among the two requests. 


\section{Resource Allocation for SPP Backup Paths}
The resource allocation problem for shared-protected backup paths assignment with \textit{MGDM} 
can be summarized as follows: \textbf{Given} a network topology, a set of traffic requests, primary paths of requests, and a reach table for different combinations of MGs and modulation formats, \textbf{decide} the modulation format, wavelength, MG and routing assignment for the backup paths for each traffic request, \textbf{constrained to} the reach and capacity of the MG combinations, the maximum number of available wavelengths, and finding backup paths for all requests, with the \textbf{objective} to either minimize total spectrum occupation or minimize the MIMO complexity.

In this paper, we propose a scalable heuristic algorithm, applicable to all four transmission scenarios introduced above.
We prioritize spectrum as the primary objective of optimization, while treating MIMO complexity as the secondary objective.
Specifically, for a new backup-path request to be accommodated (i.e., for which we need to determine routing, modulation format, wavelength, and MG), our algorithm relies on an auxiliary graph. Each auxiliary edge in the auxiliary graph denotes the possible mode group assignment in a lightpath between the end nodes of the request. Specifically, the algorithm first loops over all shortest paths between the node pair and all the wavelengths and then selects the path, wavelength, and mode groups with the minimum spectrum occupation. 
Note that if the working paths of two requests overlap in any links, their backup paths cannot share MGs in the same wavelength, and hence the spectrum occupation in the auxiliary edge is reduced. 
In the case of ties, the ones with minimum MIMO complexity are selected as secondary objectives. 


\section{Illustrative Numerical Results}

We evaluate four transmission scenarios: \textit{SMT}, \textit{MGDM}, \textit{Full-MIMO}, and  \textit{MF-MGDM} (the last three are FMT scenarios) in German topology~\cite{betker2003reference} with 17 nodes and 26 links. Each link supports 100 50-GHz dense-WDM channels. Since \textit{MF-MGDM} has a very short transmission distance (shorter than 21 km in the considered implementation~\cite{gatto2022mode,zhang2023resource}), we consider both a short-link-length scenario where all FMT approaches are applicable and a long-link-length scenario where \textit{MF-MGDM} can not be used. Specifically, we scale the maximum link length of German topology to 3 km and 380 km for the short-link-length scenario and the long-link-length scenario, respectively. 
For each scenario, we evaluate minimizing spectrum occupation as the primary objective (the cases for short-link-length scenario and long-link-length scenario are named as \textit{S} and \textit{L}, respectively). 
The data rate of requests is randomly selected from 100 Gb/s, 200 Gb/s, and 300 Gb/s. 
We evaluate the spectrum occupation and MIMO computing resources, for backup paths, of these four FMT methods with the primary optimization goal of minimizing spectrum occupation. 
We tested different traffic loads to obtain more generalizable results. Specifically, we defined traffic at 1\% rejection using DPP with \textit{SMT} as high load, and 50\% of traffic at 1\% rejection with  \textit{SMT}'s DPP as low load. 

Fig.~\ref{fig:complexity and spectrum} (a) reports the average additional spectrum occupation, used to support backup lightpaths, for the four transmission scenarios. As expected, for all scenarios,  DPP occupies more spectrum due to its inability to share among MGs. Focusing on SPP, \textit{MGDM} consistently occupies the least amount of additional backup spectrum in all scenarios. Compared to \textit{Full MIMO} and \textit{MF-MGDM}, \textit{MGDM} has reduced this metric from 20\% up to 50\%. This is due to the fact that there are more opportunities for spectrum sharing in \textit{MGDM}, as, on one side, \textit{MGDM} can share more MGs than \textit{MF-MGDM}, and, on the other side, \textit{MGDM} does not need to share all the MGs together as in \textit{FUll-MIMO}. This effect is also seen when comparing \textit{MGDM} to \textit{SMT}, since \textit{SMT} only has one mode, hence it is more difficult to share capacity of backup wavelengths. \textit{MF-MGDM} allows for limited MG sharing, hence it performs better in this metric than \textit{SMT} and \textit{Full MIMO}, but is inferior to \textit{MGDM}.

Fig.~\ref{fig:complexity and spectrum}(b) reports the average additional MIMO computing resources, used to support backup lightpaths, per Tb/s. First of all, please note that the average MIMO complexity of \textit{full MIMO} is consistently at $0$, indicating no additional MIMO resources are required for path protection. This result stems from the extended reach distance of \textit{Full MIMO}, thanks to which backup paths do not need additional MIMO with respect to their corresponding working paths.
\textit{MGDM} requires remarkably higher backup MIMO resources compared to other \textit{MF-MGDM} and MIMO methods, as it tends to select several MGs within the same wavelength to save spectrum resources (instead of establishing new lightpaths where MG with smaller complexity could be assigned), consequently consuming additional MIMO computing resources. Note though that with SPP significant savings on MIMO computing resources can be achieved with respect to DPP, showing that sharing can be effectivelty applied  to both spectrum and MIMO resources.
For \textit{MGDM}, note that the additional MIMO computing resources for high loads are lower than in case of short links for SPP, and higher for DPP. This is due to the fact that for high load opportunities of sharing increase a lot, while for DPP, at high loads, backp paths tend to be much longer than working paths. \textit{SMT} does not exhibit this phenomenon, as it involves only one mode.  

\section{Conclusion}

This work investigates the MIMO complexity and spectrum occupation of different FMT approaches with DPP and SPP in networks of various scales by proposing a scalable heuristic algorithm. The results demonstrate that although \textit{MGDM} incurs higher MIMO complexity, it also enables superior backup- shaing and hence much lower spectrum occupation, reducing it by up to 50\% compared to \textit{SMT} and \textit{Full MIMO}.

\bibliographystyle{abbrv}
\bibliography{refs}

\end{document}